\documentclass[twocolumn,showpacs,preprintnumbers]{revtex4}
\usepackage{amsmath}
\usepackage{graphicx}
\usepackage{dcolumn}
\usepackage{bm}
\usepackage{epsfig}

\begin{document}

\title{Analytic Formulae for the Matrix Elements of the Transition Operators
in the Symplectic Extension of the Interacting Vector Boson Model}
\author{H. G. Ganev, A. I. Georgieva, and V.P.Garistov}

\begin{abstract}
The tensor properties of all the generators of Sp(12,R) - the group of
dynamical symmetry of the Interacting Vector Boson Model (IVBM), are given
with respect to the reduction chain Sp(12,R) $\supset $ U(6) $\supset $ U(3) 
$\times $ U(2) $\supset $ O(3) $\times $ (U(1) $\times $ U(1)). Matrix
elements of the basic building blocks of the model are evaluated in symmetry
adapted basis along the considered chain. As a result of this, the analytic
form of the matrix elements of any operator in the enveloping algebra of the
Sp(12,R), defining a certain transition operator, can be calculated. The
procedure allows further applications of the symplectic IVBM for the
description of transition probabilities between nuclear collective states.

03.65.Fd,02.20.Sv,21.60.Fw
\end{abstract}

\maketitle
\affiliation{Institute for Nuclear Research and Nuclear Energy\\
Bulgarian Academy of Sciences, Sofia 1784, Bulgaria}


\section[]{Introduction}

In the algebraic models the use of the dynamical symmetries defined by a
certain reduction chain of the group of dynamical symmetry, yields exact
solutions for the eigenvalues and eigenfunctions of the model Hamiltonian,
which is constructed from the invariant operators of the subgroups in the
chain.

Something more, it is very simple and straightforward to calculate matrix
elements of transition operators between the eigenstates of the Hamiltonian,
as both - the basis states and the operators, can be defined as tensor
operators in respect to the considered dynamical symmetry. Then the
calculation of matrix elements is simplified by the use of the respective
generalization of the Wigner-Eckart theorem. By definition such matrix
elements give the transition probabilities between the collective states
attributed to the basis states of the Hamiltonian. The comparison of the
experimental data with the calculated transition probabilities is one of the
best tests of the validity of the considered algebraic model. With the aim
of such applications of the rotational limit of symplectic extension of
IVBM, we develop in this paper a practical mathematical approach for
explicit evaluation of the matrix elements of transitional operators in the
model.

The algebraic IVBM was developed \cite{IVBM} initially for the description
of the low lying bands of the well deformed even-even nuclei \cite{IVBMrl}.
Recently this approach was adapted to incorporate the newly observed higher
collective states, both in the first positive and negative parity bands \cite%
{Sp12U6} by considering the basis states as \textquotedblright
yrast\textquotedblright\ states for the different values of the number of
bosons $N,$ that built them. This was achieved by extending the dynamical
symmetry group $U(6)$ to the noncompact $Sp(12,R).$ The excellent results
obtained for the energy spectrum require a further investigation of the
transition probabilities in the framework of the generalized IVBM with $%
Sp(12,R)$ as a group of dynamical symmetry. Thus in the present work we
consider the tensor properties of the algebra generators (Section 2.) in
respect to the reduction chain: 
\begin{equation}
Sp(12,R)\supset U(6)\supset U(3)\times U(2)\supset O(3)\times U(1).
\label{RotLimit}
\end{equation}%
and also classify the basis states (Section 3.) by the quantum numbers \
corresponding to the \ irreducible representations of its subgroups. In this
way we are able to define the transition operators between the basis states
and then to evaluate analytically their matrix elements (Section 4.).

\section{Tensor properties of the generators of the Sp(12,R) group}

The basic building blocks of the IVBM \cite{IVBM} are the creation and
annihilation operators of the vector bosons $u_{m}^{+}(\alpha )$ and $%
u_{m}(\alpha )$ $(m=0,\pm 1;\alpha =\pm \frac{1}{2}),$ which can be
considered as components of a $6-$dimensional vector, which transform
according to the fundamental $U(6)$ irreducible representations $%
[1,0,0,0,0,0]_{6}\equiv \lbrack 1]_{6}$ and its conjugated $%
[0,0,0,0,0,-1]_{6}\equiv \lbrack 1]_{6}^{\ast }$, respectively. These
irreducible representations become reducible along the chain of subgroups (%
\ref{RotLimit}) defining the dynamical symmetry of the rotational limit of
the model \cite{IVBMrl}. This means that along with the quantum number
characterizing the representations of $U(6)$, the operators \ are also
characterized by the quantum numbers of the subgroups of chain (\ref%
{RotLimit}).

The only possible representation of the direct product of $U(3)\times U(2)$
belonging to the representation $[1]_{6}$ of $U(6)$ is $[1]_{3}.[1]_{2}$,
i.e. $[1]_{6}=[1]_{3}.[1]_{2}$. According to the reduction rules for the
decomposition $U(3)\supset O(3)$ the representation $[1]_{3}$ of $U(3)$
contains the representation $(1)_{3}$ of the group $O(3)$ giving the angular
momentum of the bosons $l=1$ with a projection $m=0,\pm 1$. The
representation $[1]_{2}$ of $U(2)$ defines the \textquotedblright
pseudospin\textquotedblright\ of the bosons $T=\frac{1}{2}$, whose
projection is given by the corresponding representation of $U(1)$, i. e. $%
\alpha =\pm \frac{1}{2}$. In this way the creation and annihilation
operators $u_{m}^{+}(\alpha )$ and $u_{m}(\alpha )$are defined as
irreducible tensors along the chain (\ref{RotLimit}) and the used phase
convention and commutation relations are the following \cite{Alisauskas}: 
\begin{equation}
\left( u_{[1]_{3}[1]_{2}m\alpha }^{[1]_{6}}\right) ^{+}=u_{[1]_{3}^{\ast
}[1]_{2}^{\ast }}^{[1]_{6}^{\ast }\quad m\alpha }=(-1)^{m+\frac{1}{2}-\alpha
}u_{[1]_{3}^{\ast }[1]_{2}^{\ast }-m-\alpha }^{[1]_{6}^{\ast }\quad }
\label{bocrao}
\end{equation}

\begin{equation*}
\left[ u_{[1]_{3}^{\ast }[1]_{2}^{\ast }}^{[1]_{6}^{\ast }\quad m\alpha
},u_{[1]_{3}[1]_{2}n\beta }^{[1]_{6}}\right] =\delta _{m,n}\delta _{\alpha
,\beta }
\end{equation*}

Initially the generators of the symplectic group $Sp(12,R)$ were written as
double tensors \cite{AGSF} with respect to $O(3)\supset O(2)$ and $%
U(2)\supset U(1)$ reductions 
\begin{equation}
A_{TT_{0}}^{LM}=\sum_{m,n}\sum_{\alpha ,\beta }C_{1m1n}^{LM}C_{\frac{1}{2}%
\alpha \frac{1}{2}\beta }^{TT_{0}}\quad u_{[1]_{3}[1]_{2}m\alpha
}^{[1]_{6}}u_{[1]_{3}^{\ast }[1]_{2}^{\ast }}^{[1]_{6}^{\ast }\quad \beta n},
\label{A}
\end{equation}%
\begin{equation}
F_{TT_{0}}^{LM}=\sum_{m,n}\sum_{\alpha ,\beta }C_{1m1n}^{LM}C_{\frac{1}{2}%
\alpha \frac{1}{2}\beta }^{TT_{0}}\quad u_{[1]_{3}[1]_{2}m\alpha
}^{[1]_{6}}u_{[1]_{3}[1]_{2}n\beta }^{[1]_{6}},  \label{F}
\end{equation}%
\begin{equation}
G_{TT_{0}}^{LM}=\sum_{m,n}\sum_{\alpha ,\beta }C_{1m1n}^{LM}C_{\frac{1}{2}%
\alpha \frac{1}{2}\beta }^{TT_{0}}\quad u_{[1]_{3}^{\ast }[1]_{2}^{\ast
}}^{[1]_{6}^{\ast }\quad \alpha m}u_{[1]_{3}^{\ast }[1]_{2}^{\ast
}}^{[1]_{6}^{\ast }\quad \beta n}.  \label{G}
\end{equation}%
Further they can be defined as irreducible tensor operators according to the
whole chain (\ref{RotLimit})\ of subgroups and expressed in terms of (\ref{A}%
), (\ref{F}) and (\ref{G}) 
\begin{equation}
A_{[\lambda ]_{3}[2T]_{2}\quad TT_{0}}^{\quad \lbrack \chi ]_{6}\quad \quad
LM}=C_{[1]_{3}[1]_{2}[1]_{3}^{\ast }[1]_{2}^{\ast }\quad \lbrack \lambda
]_{3}[2T]_{2}}^{[1]_{6}\quad \quad \lbrack 1]_{6}^{\ast }\quad \quad \quad
\lbrack \chi ]_{6}}C_{(1)_{3}(1)_{3}(L)_{3}}^{[1]_{3}[1]_{3}^{\ast }[\lambda
]_{3}}A_{TT_{0}}^{LM},  \label{Aten}
\end{equation}%
\begin{equation}
F_{[\lambda ]_{3}[2T]_{2}\quad TT_{0}}^{\quad \lbrack \chi ]_{6}\quad \quad
LM}=C_{[1]_{3}[1]_{2}[1]_{3}[1]_{2}\quad \lbrack \lambda
]_{3}[2T]_{2}}^{[1]_{6}\quad \quad \lbrack 1]_{6}\quad \quad \quad \lbrack
\chi ]_{6}}C_{(1)_{3}(1)_{3}(L)_{3}}^{[1]_{3}[1]_{3}[\lambda
]_{3}}F_{TT_{0}}^{LM},  \label{Ften}
\end{equation}%
\begin{equation}
G_{[\lambda ]_{3}[2T]_{2}\quad TT_{0}}^{\quad \lbrack \chi ]_{6}\quad \quad
LM}=C_{[1]_{3}^{\ast }[1]_{2}^{\ast }[1]_{3}^{\ast }[1]_{2}^{\ast }\quad
\lbrack \lambda ]_{3}[2T]_{2}}^{[1]_{6}^{\ast }\quad \quad \lbrack
1]_{6}^{\ast }\quad \quad \quad \lbrack \chi
]_{6}}C_{(1)_{3}(1)_{3}(L)_{3}}^{[1]_{3}^{\ast }[1]_{3}^{\ast }[\lambda
]_{3}}G_{TT_{0}}^{LM},  \label{Gten}
\end{equation}%
where, according to the lemma of Racah \cite{Racah}, the Clebsch-Gordan
coefficients along the chain are factorized by means of isoscalar factors
(IF), defined for each step of decomposition (\ref{RotLimit}). It should be
pointed out \cite{Alisauskas} that the $U(6)-$ $C_{[\lambda
_{1}]_{3}[2T_{1}]_{2}\text{ \ }[\lambda _{2}]_{3}[2T_{2}]_{2}\quad \lbrack
\lambda ]_{3}[2T]_{2}}^{[\chi _{1}]_{6}\quad \quad \text{\ \ \ \ \ \ \ \ \ \ 
}[\chi _{2}]_{6}\quad \quad \quad \lbrack \chi ]_{6}}$ and $%
U(3)-C_{(l_{1})_{3}\text{ \ \ }(l_{2})_{3}\text{\ }(L)_{3}}^{[\lambda
_{1}]_{3}\text{ }[\lambda _{2}]_{3}\text{\ }[\lambda ]_{3}}$ \ IF's,
entering in (\ref{Aten}), (\ref{Ften}) and (\ref{Gten}), are equal to $\pm $%
1 and their values are taken into account in what follows.

The tensors (\ref{Aten}), transform according to the direct product $[\chi
]_{6}$ of the corresponding $U(6)$ representations $[1]_{6}$ and $%
[1]_{6}^{\ast }$ \cite{Alisauskas}, namely 
\begin{equation}
\lbrack 1]_{6}\times \lbrack 1]_{6}^{\ast }=[1,-1]_{6}+[0]_{6},
\label{prU6A}
\end{equation}%
where $[1,-1]_{6}=[2,1,1,1,1,0]_{6}$ and $[0]_{6}=[1,1,1,1,1,1]_{6}$ is the
scalar $U(6)$ representation. Further we multiply the two conjugated
fundamental representations of $U(3)\times U(2)$%
\begin{eqnarray}
&&[1]_{3}[1]_{2}\times \lbrack 1]_{3}^{\ast }[1]_{2}^{\ast }  \notag \\
&=&([1]_{3}\times \lbrack 1]_{3}^{\ast })([1]_{2}\times \lbrack 1]_{2}^{\ast
})  \notag \\
&=&([210]_{3}\oplus \lbrack 1,1,1]_{3})\times ([2,0]_{2}\oplus \lbrack
1,1]_{2})  \notag \\
&=&[210]_{3}[2]_{2}\oplus \lbrack 210]_{3}[0]_{2}\oplus \lbrack
0]_{3}[2]_{2}\oplus \lbrack 0]_{3}[0]_{2}.  \label{u3u2prd}
\end{eqnarray}%
Obviously the first three $U(3)\times U(2)$ irreducible representations
(irreps) in the resulting decomposition (\ref{u3u2prd}) belong to the $%
[1,-1]_{6}$ of $U(6)$ and the last one to $[0]_{6}$. Introducing the
notations $u_{i}^{+}(\frac{1}{2})=p_{i}^{+}$ and $u_{i}^{+}(-\frac{1}{2}%
)=n_{i}^{+}$, the scalar operator 
\begin{equation}
A_{[0]_{3}[0]_{2}\quad 00}^{[0]_{6}\quad \quad 00}=\widehat{N}=\frac{1}{%
\sqrt{2}}\sum_{m}C_{1m1-m}^{00}\left( p_{m}^{+}p_{-m}+n_{m}^{+}n_{-m}\right)
\label{Nsc}
\end{equation}%
has the physical meaning of the total number of bosons operator $\widehat{N}%
= $ $\widehat{N_{p}}+$ $\widehat{N_{n}},$ where $\widehat{N_{p}}=\sum $ $%
p_{m}^{+}p_{m}$ , $\widehat{N_{n}}=\sum $ $n_{m}^{+}n_{m}$ and is obviously
the first order invariant of all the unitary groups $U(6),U(3)$ and $U(2).$
Hence it reduces them to their respective unimodular subgroups $SU(6),SU(3)$
and $SU(2).$ Something more, the invariant operator $(-1)^{N}$ , decomposes
the action space \textrm{H }of the $sp(12,R)$ generators to the even \textrm{%
H}$_{+}$ with $N=0,2,4,...,$ and odd \textrm{H}$_{\_}$ \ with $N=1,3,5,...,$
subspaces of the boson representations of $Sp(12,R)$ \cite{Sp2NRbr}.

The $U(3)$ irreps $[\lambda ]_{3}$ are shorthand notations of $%
[n_{1},n_{2},n_{3}]_{3}$ and are expressed in terms of Elliott's notations 
\cite{Elliott} $\left( \lambda ,\mu \right) $ with $\lambda =n_{1}-n_{2},\mu
=n_{2}-n_{3}$, so in (\ref{u3u2prd}) we have $[210]_{3}=(1,1)$ and $%
[0]_{3}=(0,0).$The corresponding values of $L$ from the $SU(3)\supset O(3)$
reduction rules are $L=1,2$ in the $(1,1)$ irrep and $L=0$ in the $(0,0)$.
The values of $T$ are $1$ and $0$ for the $U(2)$ irreps $[2]_{2}$ and $%
[0]_{2}$ respectively. Hence, the $U(2)$ irreps in the direct product
distinguish the equivalent $U(3)$ irreps that appear in this reduction and
there is not degeneracy. The tensors with $T=0$ correspond to the $SU(3)$
generators

\begin{equation}
A_{[210]_{3}[0]_{2}\quad 00}^{[1-1]_{6}\quad \ \ 1M}=\frac{1}{\sqrt{2}}%
\sum_{m,k}C_{1m1k}^{1M}\left( p_{m}^{+}p_{k}+n_{m}^{+}n_{k}\right)
\label{Lten}
\end{equation}%
\begin{equation}
A_{[210]_{3}[0]_{2}\quad 00}^{[1-1]_{6}\quad \quad 2M}=\frac{1}{\sqrt{2}}%
\sum_{m,k}C_{1m1k}^{2M}\left( p_{m}^{+}p_{k}+n_{m}^{+}n_{k}\right)
\label{Qaten}
\end{equation}%
representing the components of the angular $L_{M}$ (\ref{Lten}) and
Elliott's quadrupole $Q_{M}$ momentum (\ref{Qaten}) operators.

The tensors 
\begin{eqnarray}
A_{[0]_{3}[2]_{2}\quad 11}^{[1-1]_{6}\quad 00} &=&\sqrt{\frac{3}{2}}%
\sum_{m}p_{m}^{+}n_{-m}\sim T_{1},\qquad  \notag \\
A_{[0]_{3}[2]_{2}\quad 1-1}^{[1-1]_{6}\quad 00} &=&-\sqrt{\frac{3}{2}}%
\sum_{m}n_{m}^{+}p_{-m}\sim T_{-1}  \label{Tten}
\end{eqnarray}%
\begin{equation*}
A_{[0]_{3}[2]_{2}\quad 10}^{[1-1]_{6}\quad 00}=-\frac{1}{2}\sqrt{3}%
\sum_{m}(p_{m}^{+}p_{-m}-n_{m}^{+}n_{-m})\sim T_{0},
\end{equation*}%
correspond to the $SU(2)$ generators, which are the components of the
pseudospin operator $\widehat{T}$. And finally the tensors 
\begin{eqnarray}
A_{[210]_{3}[2]_{2}\quad 11}^{\quad \lbrack 1-1]_{6}\quad LM}
&=&\sum_{m,k}C_{1m1k}^{LM}p_{m}^{+}n_{k},\qquad  \label{A61} \\
A_{[210]_{3}[2]_{2}\quad 1-1}^{\quad \lbrack 1-1]_{6}\quad LM}
&=&\sum_{m,k}C_{1m1k}^{LM}n_{m}^{+}p_{k}  \label{A6-1}
\end{eqnarray}%
and 
\begin{equation}
A_{[210]_{3}[2]_{2}\quad 10}^{\quad \lbrack 1-1]_{6}\quad LM}=\frac{1}{\sqrt{%
2}}\sum_{m,k}C_{1m1k}^{LM}(p_{m}^{+}p_{k}-n_{m}^{+}n_{k}),  \label{A60}
\end{equation}%
with $L=1,2$ and $M=-L,-L+1,...,L$ \ extend the $U(3)\times U(2)$ algebra to
the $U(6)$ one.

By analogy, the tensors (\ref{Ften}) and (\ref{Gten}) transform according to 
\cite{Alisauskas}

\begin{equation}
\lbrack 1]_{6}\times \lbrack 1]_{6}=[2]_{6}+[1,1]_{6},  \label{prU6f}
\end{equation}
and

\begin{equation*}
\lbrack 1]_{6}^{\ast }\times \lbrack 1]_{6}^{\ast }=[-2]_{6}+[-1,-1]_{6},
\end{equation*}%
respectively. But, since the basis states of the IVBM are fully symmetric,
we consider only the fully symmetric $\ U(6)$ representations $[2]_{6}$ and $%
[-2]_{6}$ of the operators (\ref{Ften}) and (\ref{Gten}). Since the two
operators $F$ and $G$ are conjugated, i. e. $(F_{[\lambda
]_{3}[2T]_{2}TT_{0}}^{\quad \lbrack \chi ]_{6}\quad LM})^{+}$ $%
=(-1)^{\lambda +\mu +L-M+T-T_{0}}$ $G_{[\lambda ]_{3}^{\ast }[2T]_{2}^{\ast
}T-T_{0}}^{\quad \lbrack \chi ]_{6}^{\ast }\quad L-M}$, where $[\lambda
]_{3}=(\lambda ,\mu )$ we are going to present the next decompositions only
for the $F$ tensors (\ref{prU6f}). According to the decomposition rules for
the fully symmetric $U(6)$ irreps \cite{Alisauskas} we have

\begin{equation}
\lbrack 2]_{6}=[2]_{3}[2]_{2}+[1,1]_{3}[0]_{2}=(2,0)[2]_{2}+(0,1)[0]_{2}
\label{deU6fsr}
\end{equation}%
which further contain in $(2,0)$ $L=0,2$ with $T=1$ and in $(0,1)$ - $L=1$
with $T=0.$ Their explicit expressions in terms of the creation $%
p_{i}^{+},n_{i}^{+}$ and annihilation operators $p_{i},n_{i}$ at $i=0,\pm 1$
are: 
\begin{eqnarray}
F_{[2]_{3}[2]_{2}\quad 11}^{\quad \lbrack 2]_{6}\quad LM}
&=&\sum_{m,k}C_{1m1k}^{LM}p_{m}^{+}p_{k}^{+},\qquad  \notag \\
F_{[2]_{3}[2]_{2}\quad 1-1}^{\quad \lbrack 2]_{6}\quad LM}
&=&\sum_{m,k}C_{1m1k}^{LM}n_{m}^{+}n_{k}^{+}  \label{F1ex}
\end{eqnarray}%
\begin{equation*}
F_{[2]_{3}[2]_{2}\quad 10}^{\quad \lbrack 2]_{6}\quad LM}=\frac{1}{\sqrt{2}}%
\sum_{m,k}C_{1m1k}^{LM}(p_{m}^{+}n_{k}^{+}-n_{m}^{+}p_{k}^{+}),
\end{equation*}%
\begin{equation}
F_{[1,1]_{3}[0]_{2}\quad 00}^{\quad \lbrack 2]_{6}\quad \quad LM}=\frac{1}{%
\sqrt{2}}\sum_{m,k}C_{1m1k}^{LM}\left(
p_{m}^{+}n_{k}^{+}+n_{m}^{+}p_{k}^{+}\right)  \label{F0ex}
\end{equation}%
The above operators and their conjugated ones $G_{[\lambda ]_{3}^{\ast
}[2T]_{2}^{\ast }\quad TT_{0}}^{\quad \lbrack \chi ]_{6}^{\ast }\quad \quad
LM}$ \ change the number of bosons by two and realize the symplectic
extension of the $U(6)$ algebra. In this way we have listed all the
irreducible tensor operators in respect to the reduction chain (\ref%
{RotLimit}), that correspond to the infinitesimal operators of the $Sp(12,R)$
algebra.

Next we can introduce the tensor products

\begin{eqnarray}
&&T_{\qquad \qquad \lbrack \lambda ]_{3}[2T]_{2}\quad \quad \quad
TT_{0}}^{([\chi _{1}]_{6}[\chi _{2}]_{6})\quad \omega \lbrack \chi
]_{6}\quad \quad LM}=  \notag \\
&&  \notag \\
&&\sum T_{[\lambda _{1}]_{3}[2T_{1}]_{2}\quad T_{1}(T_{0})_{1}}^{[\chi
_{1}]_{6}\quad \quad \quad L_{1}M_{1}}T_{[\lambda _{2}]_{3}[2T_{2}]_{2}\quad
T_{2}(T_{0})_{2}}^{[\chi _{2}]_{6}\quad \quad \quad L_{2}M_{2}}\times  \notag
\\
&&  \label{tpr} \\
&&C_{[\lambda _{1}]_{3}[T_{1}]_{2}[\lambda _{2}]_{3}[T_{2}]_{2}\quad \lbrack
\lambda ]_{3}[2T]_{2}}^{[\chi _{1}]_{6}\quad \quad \lbrack \chi
_{2}]_{6}\quad \quad \quad \omega \lbrack \chi
]_{6}}C_{(L_{1})_{3}(L_{2})_{3}(L)_{3}}^{[\lambda _{1}]_{3}[\lambda
_{2}]_{3}[\lambda ]_{3}}\times  \notag \\
&&  \notag \\
&&C_{M_{1}\quad M_{2}\quad M}^{L_{1}\quad L_{2}\quad \quad
L}C_{(T_{0})_{1}(T_{0})_{2}\quad T_{0}}^{T_{1}\qquad T_{2}\qquad T}  \notag
\end{eqnarray}%
of two tensor operators $T_{[\lambda ]_{3}[2T]_{2}\quad TT_{0}}^{\quad
\lbrack \chi ]_{6}\quad \quad LM},$ which are as well tensors in respect to
the considered reduction chain. We use (\ref{tpr}) to obtain the tensor
properties of the operators in the enveloping algebra of $Sp(12,R),$
containing the products of the algebra generators. The products of second
degree enter the two-body interaction of the phenomenological Hamiltonian of
this limit \cite{Alisauskas}. In this particular case we are interested in
the transition operators between states differing by four bosons $%
T_{[\lambda ]_{3}[2T]_{2}\quad TT_{0}}^{[4]_{6}\quad \quad LM}$, expressed
in terms of the products of two operators $F_{[\lambda ]_{3}[2T]_{2}\quad
TT_{0}}^{\quad \lbrack 2]_{6}\quad \quad LM}$. Making use of the
decomposition (\ref{deU6fsr}) \ and the reduction rules in the chain (\ref%
{RotLimit}), we list in Table \ref{T1} all the representations of the chain
subgroups that define the transformation properties of the resulting tensors.

\begin{table}[hb]
\caption{Tensor products of two raising operators.}
\label{T1}
\smallskip \centering
\begin{tabular}{|l|l|l|l|l|l|}
\hline
\begin{tabular}{l}
$\lbrack 2]_{6}$ \\ 
$\lbrack \lambda _{1}]_{3}[2T_{1}]_{2}$%
\end{tabular}
& 
\begin{tabular}{l}
$\lbrack 2]_{6}$ \\ 
$\lbrack \lambda _{2}]_{3}[2T_{1}]_{2}$%
\end{tabular}
& 
\begin{tabular}{l}
$\lbrack 4]_{6}$ \\ 
$\lbrack \lambda ]_{3}[2T]_{2}$%
\end{tabular}
& 
\begin{tabular}{l}
$O(3)$ \\ 
K; $L$%
\end{tabular}
& 
\begin{tabular}{l}
$U(2)$ \\ 
$T$%
\end{tabular}
& 
\begin{tabular}{l}
$U(1)$ \\ 
$T_{0}$%
\end{tabular}
\\ \hline
$(2,0)[2]_{2}$ & $(2,0)[2]_{2}$ & $(4,0)[4]_{2}$ & 0;$0,2,4$ & $2$ & $0,\pm
1,\pm 2$ \\ \hline
$(2,0)[2]_{2}$ & $(0,1)[0]_{2}$ & $(2,1)[2]_{2}$ & 1;$1,2,3$ & $1$ & $0,\pm
1 $ \\ \hline
$(0,1)[0]_{2}$ & $(0,1)[0]_{2}$ & $(0,2)[0]_{2}$ & 0;$0,2$ & $0$ & $0$ \\ 
\hline
\end{tabular}%
\end{table}

The basis states in the $\mathcal{H}_{+}$ are also obtained as symmetrized
tensor products of different degrees of the tensor operators $F_{[\lambda
]_{3}[2T]_{2}\quad TT_{0}}^{\quad \lbrack 2]_{6}\quad \quad LM}$.

\section{Construction of the symplectic basis states of the IVBM}

In order to clarify the role of the tensor operators introduced in the
previous section as transition operators and to simplify the calculation of
their matrix elements, the basis for the Hilbert space must be symmetry
adapted to the \ algebraic structure along the considered subgroup chain (%
\ref{RotLimit}). It is evident from (\ref{F1ex}) and (\ref{F0ex}), that the
basis states of the IVBM in the $\mathcal{H}_{+}$ \ ($N-$even) \ subspace of
the boson representations of $\ Sp(12,R)$ \ can be obtained by a consecutive
application of the raising operators \ $F_{[\lambda ]_{3}[2T]_{2}\quad
TT_{0}}^{\quad \lbrack 2]_{6}\quad \quad LM}$ \ on the boson vacuum (ground)
state $\mid 0\quad \rangle ,$ annihilated by the tensor operators $%
G_{[\lambda ]_{3}[2T]_{2}\quad TT_{0}}^{\quad \lbrack \chi ]_{6}\quad \quad
LM}$ $\mid 0\quad \rangle =0$ and $A_{[\lambda ]_{3}[2T]_{2}\quad
TT_{0}}^{\quad \lbrack \chi ]_{6}\quad \quad LM}\mid 0\quad \rangle =0$.

Thus, in general a basis for the considered dynamical symmetry of the IVBM
can be constructed by applying the multiple symmetric coupling (\ref{tpr})
of the raising tensors $F_{[\lambda _{i}]_{3}[2T_{i}]_{2}\quad
T_{i}T_{0i}}^{\quad \lbrack 2]_{6}\quad \quad L_{i}M_{i}}$\ with itself \ - $%
[F\times \ldots \times \quad F]_{[\lambda ]_{3}[2T]_{2}\quad TT_{0}}^{\quad
\lbrack \chi ]_{6}\quad \quad LM}$ . Note that only fully symmetric tensor
products $[\chi ]_{6}\equiv \lbrack N]_{6}$ \ are nonzero, since the raising
operator commutes with itself. The possible $U(3)$ couplings are enumerated
by the set $[\lambda ]_{3}=\{[n_{1},n_{2},0]\equiv
(n_{1}-n_{2},n_{2},);n_{1}\geq n_{2}\geq 0$ $\}$. In terms of the notations $%
(\lambda ,\mu )$ the $SU(3)$ content of the $U(6)$\ symmetric tensor $%
[N]_{6} $ is determined by $\lambda =n_{1}-n_{2},\mu =n_{2}$. The number of
copies of the operator $F$ \ in the symmetric tensor\ product $[N]_{6}$ is $%
N/2$, where $N=n_{1}+n_{2}=\lambda +2\mu $ \cite{Sp12U6}. Each raising
operator will increase the number of bosons $N$ by two. Then, the resulting
infinite basis is denoted by $|[N](\lambda ,\mu );KLM;TT_{0}\quad \rangle $,
where $KLM$ are the quantum numbers for the non-orthonormal basis of the
irrep $(\lambda ,\mu )$.

The $Sp(12,R)$ classification scheme for the $SU(3)$ boson representations
obtained by applying the reduction rules \cite{Sp12U6} for the irreps in the
chain (\ref{RotLimit}) for even value of the number of bosons $N$ \ is shown
on Table \ref{T2}. Each row (fixed $N$) of the table corresponds to a given
irreducible representation of the $U(6)$ algebra. Then the possible values
for the pseudospin, given in the column next to the respective value of $N$,
are $T=\frac{N}{2},\frac{N}{2}-1,...$ $0$. Thus when $N$ and $T$ are fixed, $%
2T+1$ equivalent representations of the group $SU(3)$ arise. Each of them is
distinguished by the eigenvalues of the operator $T_{0}:-T,-T+1,...,T,$
defining the columns of Table \ref{T2}. The same $SU(3)$ representations $%
(\lambda ,\mu )$ arise for the positive and negative eigenvalues of $T_{0}$. 
\begin{widetext}
\begin{table}[bh]
\caption{Classification of the basis states.}
\label{T2}
\smallskip \centering%
\begin{tabular}{||l||rrrr|l|}
\hline\hline
$N$ T & $T_{_{0}}\backslash $ $...\ \pm 4$ & \multicolumn{1}{||r}{$\pm 3$} & 
\multicolumn{1}{||r}{$\pm 2$} & \multicolumn{1}{||r|}{$\ \pm 1$} & 
\multicolumn{1}{||l||}{$\ \ 0$} \\ \hline\hline
\multicolumn{1}{||r||}{$0%
\begin{tabular}{l}
$0$%
\end{tabular}%
$} & \multicolumn{1}{||c}{} & \multicolumn{1}{l}{} & \multicolumn{1}{l}{} & 
\multicolumn{1}{c|}{$\swarrow F_{[2]_{3}[2]_{2}\quad }^{\quad \lbrack
2]_{6}\quad }$} & 
\begin{tabular}{l}
$(0,0)$%
\end{tabular}
\\ \cline{1-1}\cline{5-6}
\multicolumn{1}{||r||}{$2%
\begin{tabular}{l}
$1$ \\ 
$0$%
\end{tabular}%
$} & \multicolumn{1}{||c}{} & \multicolumn{1}{l}{} & \multicolumn{1}{l}{$%
F_{[1,1]_{3}[0]_{2}}^{\quad \lbrack 2]_{6}\quad }\downarrow $} & 
\multicolumn{1}{|r|}{$%
\begin{tabular}{c}
$\Longrightarrow $ \\ 
\multicolumn{1}{l}{$A_{[2,1]_{3}[0]_{2}}^{\quad \lbrack 1-1]_{6}}$}%
\end{tabular}%
$%
\begin{tabular}{l}
$(2,0)$ \\ 
\multicolumn{1}{c}{$-$}%
\end{tabular}%
} & 
\begin{tabular}{l}
$(2,0)$ \\ 
$(0,1)$%
\end{tabular}
\\ \cline{1-1}\cline{4-6}
\multicolumn{1}{||r||}{$4%
\begin{tabular}{l}
$2$ \\ 
$1$ \\ 
$0$%
\end{tabular}%
$} & \multicolumn{1}{||c}{} & \multicolumn{1}{l}{} & \multicolumn{1}{|r}{%
\begin{tabular}{l}
$(4,0)$ \\ 
\multicolumn{1}{c}{$-$} \\ 
\multicolumn{1}{c}{$-$}%
\end{tabular}%
} & \multicolumn{1}{|r|}{$A_{[2,1]_{3}[2]_{2}}^{\quad \lbrack
1-1]_{6}}\Downarrow $\ 
\begin{tabular}{l}
$(4,0)$ \\ 
$(2,1)$ \\ 
\multicolumn{1}{c}{$-$}%
\end{tabular}%
} & 
\begin{tabular}{l}
$(4,0)$ \\ 
$(2,1)$ \\ 
$(0,2)$%
\end{tabular}
\\ \cline{1-1}\cline{3-5}\cline{5-6}
\multicolumn{1}{||r||}{$6%
\begin{tabular}{l}
$3$ \\ 
$2$ \\ 
$1$ \\ 
$0$%
\end{tabular}%
$} & \multicolumn{1}{||c}{%
\begin{tabular}{l}
$A_{[0]_{3}[2]_{2}}^{\quad \lbrack 1-1]_{6}}$ \\ 
\multicolumn{1}{c}{$\rightarrow $}%
\end{tabular}%
} & \multicolumn{1}{|r}{%
\begin{tabular}{l}
$(6,0)$ \\ 
\multicolumn{1}{c}{$-$} \\ 
\multicolumn{1}{c}{$-$} \\ 
\multicolumn{1}{c}{$-$}%
\end{tabular}%
} & \multicolumn{1}{|r}{%
\begin{tabular}{l}
$(6,0)$ \\ 
$(4,1)$ \\ 
\multicolumn{1}{c}{$-$} \\ 
\multicolumn{1}{c}{$-$}%
\end{tabular}%
} & \multicolumn{1}{|r|}{%
\begin{tabular}{l}
$(6,0)$ \\ 
$(4,1)$ \\ 
$(2,2)$ \\ 
\multicolumn{1}{c}{$-$}%
\end{tabular}%
} & 
\begin{tabular}{l}
$(6,0)$ \\ 
$(4,1)$ \\ 
$(2,2)$ \\ 
$(0,3)$%
\end{tabular}
\\ \cline{1-2}\cline{2-6}
\multicolumn{1}{||r||}{$8%
\begin{tabular}{l}
$4$ \\ 
$3$ \\ 
$2$ \\ 
$1$ \\ 
$0$%
\end{tabular}%
$} & 
\begin{tabular}{l}
$(8,0)$ \\ 
\multicolumn{1}{c}{$-$} \\ 
\multicolumn{1}{c}{$-$} \\ 
\multicolumn{1}{c}{$-$} \\ 
\multicolumn{1}{c}{$-$}%
\end{tabular}
& \multicolumn{1}{|r}{%
\begin{tabular}{l}
$(8,0)$ \\ 
$(6,1)$ \\ 
\multicolumn{1}{c}{$-$} \\ 
\multicolumn{1}{c}{$-$} \\ 
\multicolumn{1}{c}{$-$}%
\end{tabular}%
} & \multicolumn{1}{|r}{$\ \ $%
\begin{tabular}{l}
$(8,0)$ \\ 
$(6,1)$ \\ 
$(4,2)$ \\ 
\multicolumn{1}{c}{$-$} \\ 
\multicolumn{1}{c}{$-$}%
\end{tabular}%
} & \multicolumn{1}{|r|}{%
\begin{tabular}{l}
$(8,0)$ \\ 
$(6,1)$ \\ 
$(4,2)$ \\ 
$(2,3)$ \\ 
\multicolumn{1}{c}{$-$}%
\end{tabular}%
} & 
\begin{tabular}{l}
$(8,0)$ \\ 
$(6,1)$ \\ 
$(4,2)$ \\ 
$(2,3)$ \\ 
$(0,4)$%
\end{tabular}
\\ \cline{1-1}\cline{2-2}\cline{2-6}
\multicolumn{1}{||r||}{$...$} & $...$ & $...$ & \multicolumn{1}{|r}{$...$} & 
\multicolumn{1}{|r|}{$...$} & \multicolumn{1}{|r|}{$...$}%
\end{tabular}%
\end{table}
\end{widetext}
Now it is clear which of the tensor operators act as transition operators
between the basis states ordered in the classification scheme presented on
Table \ref{T2}. The operators $F_{[\lambda ]_{3}[2T]_{2}\quad TT_{0}}^{\quad
\lbrack 2]_{6}\quad \quad LM}$ with $T_{0}=0$ (\ref{F0ex}) give the
transitions between two neighboring cells $(\downarrow )$ from one column,
while the ones with $T_{0}=\pm 1$(\ref{F1ex}) change the column as well $%
(\swarrow )$. The \ tensors $A_{[2,1]_{3}[0]_{2}}^{\quad \lbrack 1-1]_{6}}$ (%
\ref{Lten}) and (\ref{Qaten}), which correspond to the $SU(3)$ generators do
not change the $SU(3)$ representations $(\lambda ,\mu ),$ but can change the
angular momentum $L$ inside it $(\Longrightarrow )$. The $SU(2)$ generating
tensors $A_{[0]_{3}[2]_{2}}^{\quad \lbrack 1-1]_{6}}$(\ref{Tten}) change the
projection $T_{_{0}}(\rightarrow )$ of the pseudospin $T$ and in this way
distinguish the equivalent $SU(3)$ irreps belonging to the different columns
of the same row of Table \ref{T2}. Inside a given cell the transition
between the different $SU(3)$ irreps $(\Downarrow )$ is realized by the
operators $A_{[2,1]_{3}[2]_{2}}^{\quad \lbrack 1-1]_{6}}$ (\ref{A61}), (\ref%
{A6-1}) and (\ref{A60}), that represent the $U(6)$ generators. The action of
the tensor operators on the $SU(3)$ vectors inside a given cell or between
the cells of Table \ref{T2}. is also schematically presented on it with
corresponding arrows. In physical applications sequences of $SU(3)$ vectors
are attributed to sequences of collective states belonging to different
bands in the nuclear spectra. By means of the above analysis, the
appropriate transition operators can be defined as appropriate combinations
of the tensor operators given in Section 2.

\section{Matrix elements of the transition operators in symmetry adapted
basis}

Matrix elements of the $Sp(12,R)$ algebra can be calculated in several ways.
A direct method is to use the $Sp(12,R)$ commutation relations \cite{IVBM}
to derive recursion relations. Another is to start from approximate matrix
element and proceed by successive approximations, adjusting the matrix
elements until the commutation relations are precisely satisfied \cite%
{RRmatr}. The third method is to make use of a vector-valued coherent-state
representation theory \cite{AGSF},\cite{VCSforSp} to relate the matrix
elements to the known matrix elements of a much simpler Weyl algebra.

However, in the present paper we use another technique for calculation of
the matrix elements of the $\ Sp(12,R)$ algebra, based on the fact that the
representations of the $SU(3)$ subgroup in IVBM are build with the help of
the two kind of vector bosons, which is in some sense simpler than the
construction of the $SU(3)$ representations in IBM \cite{IBM} and $Sp(6,R)$
symplectic model\cite{Rowe}.

In the preceding sections we expressed the $Sp(12,R)$ generators $%
F_{TT_{0}}^{LM},$ $G_{TT_{0}}^{LM},$ $A_{TT_{0}}^{LM}$\ and the basis states
as components of irreducible tensors in respect to the reduction chain (\ref%
{RotLimit}). Thus, for calculating their matrix elements, we have the
advantage of using the Wigner-Eckart theorem in two steps. For the $%
SU(3)\rightarrow SO(3)$ and $SU(2)\rightarrow U(1)\times U(1)$ reduction we
need the standard $SU(2)$ Clebsch-Gordan coefficient (CGC) 
\begin{widetext}
\begin{equation}
\begin{tabular}{l}
$\langle \lbrack N^{\prime }](\lambda ^{\prime },\mu ^{\prime });K^{\prime
}L^{\prime }M^{\prime };T^{\prime }T_{0}^{\prime }|T_{[\lambda
]_{3}[2t]_{2}\quad tt_{0}}^{[\chi ]_{6}\quad \quad lm}|[N](\lambda ,\mu
);KLM;TT_{0}\quad \rangle =$ \\ 
$\langle \lbrack N^{\prime }](\lambda ^{\prime },\mu ^{\prime });K^{\prime
}L^{\prime }||T_{[\lambda ]_{3}[2t]_{2}}^{[\chi ]_{6}}||[N](\lambda ,\mu
);KL\rangle C_{LMlm}^{L^{\prime }M^{\prime }}C_{TT_{0}tt_{0}}^{T^{\prime
}T_{0}^{\prime }}.$%
\end{tabular}
\label{ME}
\end{equation}
\end{widetext}
For the calculation of the double-barred reduced matrix elements in (\ref{ME}%
) we use the form \cite{Ros90}: 
\begin{equation}
\begin{tabular}{l}
$\langle \lbrack N^{\prime }](\lambda ^{\prime },\mu ^{\prime });K^{\prime
}L^{\prime }||T_{[\lambda ]_{3}[2t]_{2}\quad tt_{0}}^{[\chi ]_{6}\quad \quad
lm}||[N](\lambda ,\mu );KL\rangle $ \\ 
$=\langle \lbrack N^{\prime }](\lambda ^{\prime },\mu ^{\prime
})|||T_{[\lambda ]_{3}[2t]_{2}}^{[\chi ]_{6}}|||[N](\lambda ,\mu )\rangle
C_{KL\quad (L)_{3}\quad K^{\prime }L^{\prime }}^{(\lambda ,\mu )[\lambda
]_{3}(\lambda ^{\prime },\mu ^{\prime })}$%
\end{tabular}
\label{3-barredME}
\end{equation}%
where $C_{KL\quad (L)_{3}\quad K^{\prime }L^{\prime }}^{(\lambda ,\mu
)[\lambda ]_{3}(\lambda ^{\prime },\mu ^{\prime })}$ is a reduced $SU(3)$\
Clebsch-Gordan coefficient (CGC), which are known analytically in many
special cases \cite{Ver}, \cite{He},\cite{RoBa},\cite{DrAk} and computer
codes are available to calculate them numerically \cite{AkDr}, \cite{BaDr},
and \cite{BaRoDr}. Hence for the evaluation of the matrix elements (\ref{ME}%
) of the $Sp(12,R)$ operators only their reduced triple-barred matrix
elements are required. In order to evaluate them we first obtain the matrix
elements of the creation $u_{m}^{+}(\alpha )$ and annihilation $u_{m}(\alpha
)$ operators $(m=$ $0,\pm 1;\alpha =\pm \frac{1}{2})$ of the vector bosons
that build them. The latter act in the Hilbert space $\mathcal{H}$ of the
boson representation of the algebra of $Sp(12,R)$ with a vacuum $|0\rangle $
defined by $u_{m}(\alpha )|0\rangle =0.$ In the notations $u_{i}^{+}(\frac{1%
}{2})=p_{i}^{+}$ and $u_{i}^{+}(-\frac{1}{2})=n_{i}^{+},$ an orthonormal
basis in $\mathcal{H}$ is introduced in the following way \cite{Sp2NRbr}

\begin{equation}
\left\vert \pi ,\nu \right\rangle =\underset{i,k=0,\pm 1}{{\Huge \Pi }}\frac{%
(p_{i}^{+})^{\pi _{i}}}{\sqrt{\pi _{i}!}}\frac{(n_{k}^{+})^{\nu _{k}}}{\sqrt{%
\nu _{k}!}}|0\rangle ,  \label{bost}
\end{equation}%
where $\pi \equiv \{\pi _{1},\pi _{0},\pi _{-1}\}$ run over the set of three
nonnegative numbers for which $N_{p}=\underset{i}{\sum }$ $\pi _{i}$ and the
same is valid for $\nu \equiv \{\nu _{1},\nu _{0},\nu _{-1}\}$ with\ $N_{n}=%
\underset{i}{\sum }$ $\nu _{i},$ where $N_{p}$ \ and $N_{n}$ give the number
of bosons of each kind and the total number $\ $of bosons, that build each
state is $N=N_{p}+N_{n}$. These numbers are eigenvalues of the corresponding
operators $\widehat{N_{p}}=\sum $ $p_{m}^{+}p_{m}$ , $\widehat{N_{n}}=\sum $ 
$n_{m}^{+}n_{m}$ and $\widehat{N}=$ $\widehat{N_{p}}+$ $\widehat{N_{n}}$ (%
\ref{Nsc}):

\begin{equation}
\widehat{N_{p}}\left\vert \pi ,\nu \right\rangle =N_{p}\left\vert \pi ,\nu
\right\rangle ,\qquad \widehat{N_{n}}\left\vert \pi ,\nu \right\rangle
=N_{n}\left\vert \pi ,\nu \right\rangle  \label{evNpn}
\end{equation}%
\begin{equation}
\widehat{N}\left\vert \pi ,\nu \right\rangle =N\left\vert \pi ,\nu
\right\rangle  \label{evN}
\end{equation}%
As a result of the above relations the basis states $\left\vert \pi ,\nu
\right\rangle $ can be labeled with the quantum numbers $N,N_{p},N_{n}.$ In
the considered reduction chain (\ref{RotLimit}) the labels of the $%
SU(3)-(\lambda ,\mu )$ irreps are related to the numbers of the introduced $%
n $ and $p$ -vector bosons in the following way $\lambda =N_{p}-N_{n},\mu
=N_{n},N=\lambda +2\mu \ \ $\cite{Sp12U6}. It is simple to see that the
eigenvalues in (\ref{evNpn}) $N_{p}\equiv n_{1}$ and $N_{n}\equiv n_{2}$ ,
where $n_{1},n_{2}$ defined the $U(3)$ tensor properties $[\lambda ]_{3}$ in
the tensor operators (\ref{Aten}), (\ref{Ften}) and (\ref{Gten}) and the
symmetry adapted basis states, given in Table \ref{T2}. Also the eigenvalue $%
N$ in (\ref{evNpn}) corresponds to the totally symmetric $U(6)$ irrep $[N]$
that defines the tensor operators and the basis states tensor properties.
Hence the basis (\ref{bost}) can be equivalently labeled by $|[N];(\lambda
,\mu )\rangle $ with $\dim (\lambda ,\mu )=\frac{1}{2}(\lambda +\mu
+2)(\lambda +1)(\mu +1)$ included in the normalization of the states. The
action of any component of the boson creation and annihilation operators is
standardly given by%
\begin{equation}
\begin{tabular}{l}
$p_{i}^{+}|[N];(\lambda ,\mu )\rangle =\sqrt{\frac{(N+1)\dim (\lambda ,\mu )%
}{\dim (\lambda ^{\prime },\mu ^{\prime })}}|[N+1];(\lambda ^{\prime },\mu
^{\prime })\rangle $ \\ 
\\ 
$=\sqrt{\frac{(2\mu +\lambda +1)(\lambda +\mu +2)(\lambda +1)}{(\lambda +\mu
+3)(\lambda +2)}}|[N+1];(\lambda +1,\mu )\rangle ,$%
\end{tabular}
\label{p+action}
\end{equation}
\begin{equation}
\begin{tabular}{l}
$n_{i}^{+}|[N];(\lambda ,\mu )\rangle =$ \\ 
\\ 
$\sqrt{\frac{(2\mu +\lambda +1)(\lambda +\mu +2)(\lambda +1)(\mu +1)}{%
\lambda (\lambda +\mu +1)(\mu +2)}}|[N+1];(\lambda -1,\mu +1)\rangle ,$%
\end{tabular}
\label{n+action}
\end{equation}%
\begin{equation}
\begin{tabular}{l}
$p_{i}|[N];(\lambda ,\mu )\rangle =$ \\ 
\\ 
$\sqrt{\frac{(2\mu +\lambda )(\lambda +\mu +2)(\lambda +1)}{\lambda (\lambda
+\mu +1)}}|[N-1];(\lambda -1,\mu )\rangle ,$%
\end{tabular}
\label{paction}
\end{equation}%
\begin{equation}
\begin{tabular}{l}
$n_{i}|[N];(\lambda ,\mu )\rangle =$ \\ 
\\ 
$\sqrt{\frac{(2\mu +\lambda )(\lambda +1)(\mu +1)}{(\lambda +2)\mu }}%
|[N-1];(\lambda +1,\mu -1)\rangle .$%
\end{tabular}
\label{naction}
\end{equation}%
With the help of relations ( \ref{p+action})$-$(\ref{naction})\ we can
evaluate the corresponding matrix elements of the building blocks of the
IVBM. They correspond to the triple-barred reduced matrix elements in (\ref%
{3-barredME}) as they depend only on the $U(6)\rightarrow SU(3)$ quantum
numbers. \ The above expressions for the action of the creation and
annihilation operators are very simple and useful, as only a single
resulting state is obtained. The explicitly presented in Section 2. $%
Sp(12,R) $ generators, as tensor operators in terms of bilinear products of $%
p_{i}^{+},n_{i}^{+},$ $p_{i}$ and $n_{i}$ can be considered as \ coupled $%
U(6)\rightarrow SU(3)$ tensors (\ref{tpr}) and their matrix elements
calculated using the above expressions where the state resulting from the
action of the operators taken as an intermediate state \cite{Ros90}.

\section{Conclusions}

In the present paper we investigate the tensor properties of the algebra
generators of $Sp(12,R)$ with respect to the reduction chain (\ref{RotLimit}%
). $Sp(12,R)$ is the group of dynamical symmetry of the IVBM and the
considered chain of subgroups was applied \ in \cite{Sp12U6} for the
description of positive and negative parity bands in well deformed nuclei.
The basis states of the model Hamiltonian are also classified by the quantum
numbers \ corresponding to the irreducible representations of its subgroups
and in this way the symmetry adapted basis in this limit of the IVBM is
constructed. The action of the symplectic generators as transition operators
between the basis states is presented. Simple analytical expressions for the
matrix elements of the basic building blocks of the model are obtained as
well. Making use of the latter and respective generalization of the
Wigner-Eckart theorem one is able to evaluate explicitly the matrix elements
of the transition operators expressed in terms of tensor operators. By
definition such matrix elements give the transition probabilities between
the collective states attributed to the basis states of the Hamiltonian. In
this way useful mathematical tool is developed, which will allow future
applications of the symplectic IVBM in the description of different
collective features of the nuclear systems. Furthermore, we hope that such
investigations will contribute for deeper understanding of the physical
meaning of the mathematical structures of the model and more correct
evaluation of its limits of applicability.

\section*{Acknowledgments}

The help and discussion on this work of Dr. K. Sviratcheva are acknowledged.

\end{document}